\documentclass{PoS}
\usepackage{graphicx}

\def\be{\begin{equation}}
\def\ee{\end{equation}}
\def\ba{\begin{eqnarray}}
\def\ea{\end{eqnarray}}

\title{Top production at large $p_t$ at NLO+NLL accuracy}

\ShortTitle{Top production at large $p_t$ at NLO+NLL accuracy}

\author{\speaker{Matteo Cacciari}\thanks{This work has been performed with the
collaboration of Fr\'ed\'eric Dreyer and Emanuele Re.}\\
        Universit\'e Paris Diderot, Paris, France\\
	Laboratoire de Physique Th\'eorique et Hautes Energies (LPTHE), UMR 7589
	CNRS \& Sorbonne Universit\'e, 4 Place Jussieu, F-75252, Paris, France \\
        E-mail: \email{cacciari@lpthe.jussieu.fr}}

\abstract{We introduce a new version of the FONLL code, now capable of calculating
differential distributions for top quark production with next-to-leading-log
resummation of $\log(p_t/m)$ terms. Numerical results for LHC and FCC
kinematics are presented. In the transverse momentum region
presently explored by ATLAS and CMS, no significant difference with respect to 
available fixed order predictions is predicted by
FONLL. The large transverse momentum
resummation of FONLL may instead become relevant if top is ever measured at
transverse momentum scales of several TeV.}

\FullConference{XXVI International Workshop on Deep-Inelastic Scattering and Related Subjects (DIS2018)\\
		16-20 April 2018\\
		Kobe, Japan}

\begin{document}

\section{Introduction}

Heavy quark (charm, bottom and top) production in proton-proton collisions has
been measured at the LHC from the very beginning of its run in 2010. Data for
charm and bottom cross sections are available in a wide range of rapidity and
transverse momentum, and have generally been found to be in good agreement with
QCD predictions. The same is true for inclusive top cross sections, which have
been found to be correctly described by next-to-next-to-leading order (NNLO) QCD
calculations.

In recent years, the LHC has delivered an integrated luminosity sufficiently
large to allow for measurements of differential top cross sections in a regime where the
transverse momentum $p_t$ starts being significantly larger than the top mass
$m$: recent papers from ATLAS~\cite{Aaboud:2018eqg} and CMS~\cite{Sirunyan:2018wem} report measurements up to
transverse momenta of the order of 1 TeV. The question of the
performance of the theoretical predictions in this previously unexplored
regime\footnote{While one may argue that a similar regime has already been probed by
charm and bottom data, in that case a non-perturbative fragmentation function is an
unavoidable part of the theoretical prediction, and it may to some extent ``compensate''
for possible shortcomings of the perturbative calculation. This won't be the case for top
quark data.}
becomes therefore a pertinent one. Moreover, design studies for large future hadron
colliders like the FCC, where top quarks may be measured in the
multi-TeV range, could start making use of such knowledge.

Perturbative QCD predictions for transverse-momentum distributions of heavy
quarks have existed since a long time at next-to-leading order (NLO) accuracy~\cite{Nason:1989zy,Beenakker:1988bq}, and more
recently have also become available at NNLO~\cite{Czakon:2015owf,Czakon:2016ckf,Czakon:2016dgf}. These calculations
have been compared to the ATLAS and CMS data from refs.~\cite{Aaboud:2018eqg,Sirunyan:2018wem} and found in good
agreement with the experimental data.

Nevertheless, reasons exist to pursue alternative calculations. On one hand,
while transverse momenta of the order of 1 TeV for top production  are likely
not large enough to make resummation to all orders of $\log(p_t/m)$ terms 
necessary, it is worth starting to ask the question of the range of validity of
a fixed order perturbative calculation. On the other hand, the NNLO calculation
for the transverse momentum distribution is only available in numerical form,
and no results presently exist for top production at the FCC because the
numerics become challenging in this kinematical regime.

These reasons have motivated our choice of extending the FONLL~\cite{Cacciari:1998it} calculation and
code to the case of top production. FONLL has been used in the past twenty
years to evaluate charm and bottom production at very large transverse momenta,
$p_t \gg m$, resumming  $\log(p_t/m)$ terms to all orders to
next-to-leading-log (NLL) accuracy. Matching with the NLO fixed order
calculation of~\cite{Nason:1989zy} gives an overall NLO+NLL result.

While in principle straightforward, some work is needed to obtain an explicit
extension of the FONLL code to the top quark production case, because the
implementation of the calculation was originally performed for charm and bottom
quarks only, with the number of flavours hard-coded to a maximum of five. Other
necessary ingredients are an implementation of the running of the strong
coupling $\alpha_s$ with six flavours, which is easily written, and parton
distribution functions (PDFs) also evolved with six flavours and providing
distributions also for top quarks. This latter ingredient was not easily
available twenty years ago, but today many modern PDF groups routinely publish
sets with this feature.

\section{FONLL}

We recall here briefly the structure of the FONLL calculation. 

FONLL~\cite{Cacciari:1998it} matches a massive, fixed-order (NLO) calculation of heavy quark 
production~\cite{Nason:1989zy} with a massless calculation that resums to all orders (to
NLL accuracy) massive logarithms of the form $\alpha_s^n \log^k(p_t/m)$, originating from
gluon emission off massive quarks and from gluon splitting~\cite{Cacciari:1993mq}. The
matched calculation (called ``FONLL'' for historical reasons, but it could also be called
``NLO+NLL'') provides therefore all the mass terms up to order $\alpha_s^3$, but also the
resummation of the aforementioned logs, that become important in the $p_t \gg m$ region.
A direct consequence of the resummation is that the perturbative uncertainty band of
FONLL is usually much smaller than that of the NLO calculation (except, of course, in the region
$p_t \simeq m$ or below, where the matching is dominated by the NLO result), and the
FONLL 
$p_t$ distribution tends to be softer than the NLO one at large $p_t$, a direct
consequence of a larger energy loss due to multiple gluon emissions.

Schematically, the FONLL matching can be written as
\be
\mathrm{FONLL} = \mathrm{FO} + (\mathrm{RS} - \mathrm{FOM0}) G(p_t,m) \, .
\ee
In this equation, FO is the massive NLO calculation, RS is the massless resummed one, and
FOM0 is the massless limit of FO, where only $\alpha_s \log(p_t/m)$ terms are
kept. Therefore, the RS - FOM0 subtraction ensures that terms
that are present in both FO and in RS are not double-counted. Finally, 
$G(p_t,m)$ is a (to some extent) arbitrary damping function that prevents
spurious higher order (but artificially massless) terms from giving an 
unphysically large contribution. An ambiguity similar to the one related to the arbitrariness in the choice of
$G$ is present in essentially all calculations of matched type\footnote{See 
\cite{Cacciari:1998it} for a more in-depth discussion about the choice of the damping
function and about the formulation of FONLL in general.}.

The RS result can, also schematically, be written as
\be
\label{eq:rs}
\mathrm{RS} = \mathrm{PDF}_i \otimes \mathrm{PDF}_j \otimes d\sigma_{ij \to k}(p_t)
\otimes \mathrm{FF}_k \, ,
\ee
where PDF$_i$ are parton distribution functions (including the one for the
nominally heavy quark that one is interested in calculating), $d\sigma_{ij \to
k}(p_t)$ are massless partonic cross sections, and FF$_k$ are 
fragmentation functions. $i$, $j$ and $k$ are flavour indices, and they run over
all active flavours, including the heavy one. DGLAP evolution of the PDFs and 
of the FF resums the  $\alpha_s^n\log^k(p_t/m)$ terms to all orders to NLL accuracy.

\section{Results}

The new FONLL code for top has been run at LHC ($\sqrt{S} = 13$~TeV) and FCC
($\sqrt{S} = 100$~TeV) energies, and top production has been studied up to very large
transverse momenta.

\begin{figure}[t]
\includegraphics[width=0.5\textwidth]{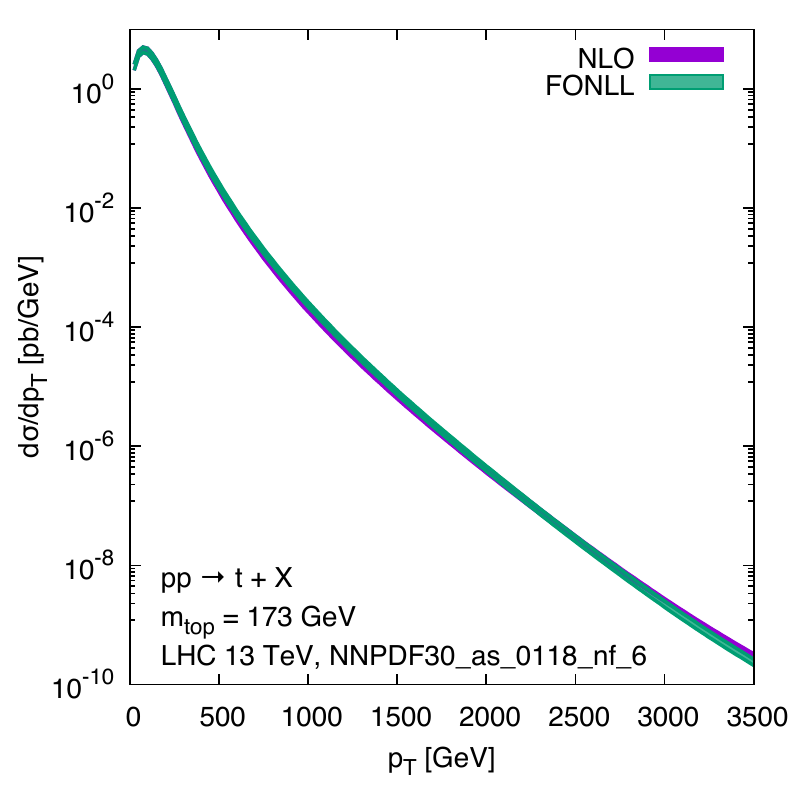}
\includegraphics[width=0.5\textwidth]{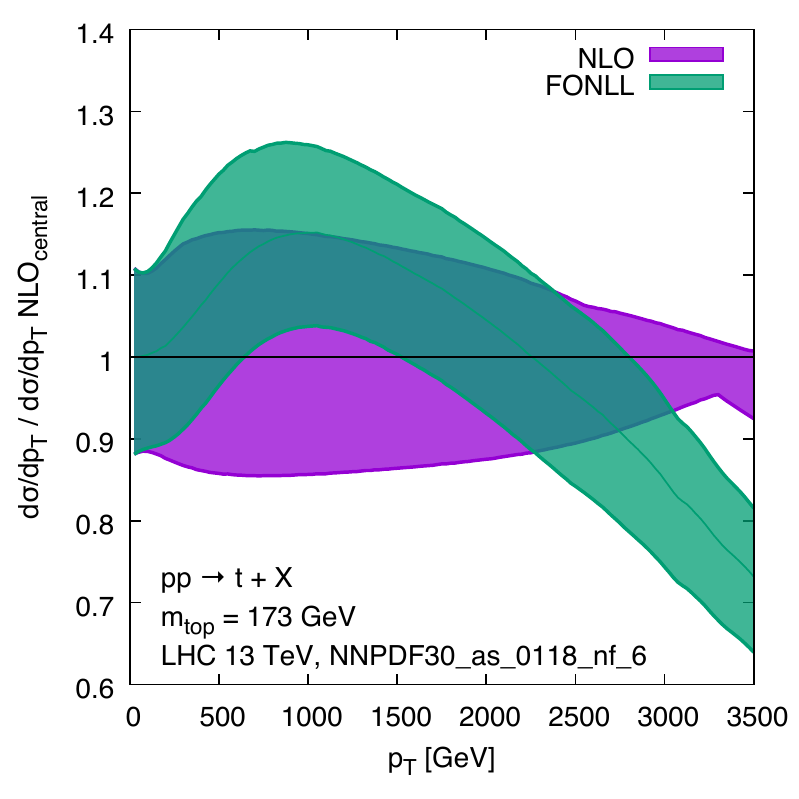}
\caption{\label{fig:lhc} Comparison of NLO and FONLL transverse momentum
distribution for top quark production at the LHC.}
\end{figure}

Figure~\ref{fig:lhc} compares the predictions for the top quark transverse
momentum distribution at the LHC at NLO and FONLL accuracy. As expected, the
perturbative 
uncertainty of the NLO prediction tends to increase with $p_t$ (the subsequent 
apparent reduction at larger $p_t$ being simply due to accidental compensations between
scale variations), while the FONLL prediction's uncertainty band is smaller and
more stable with increasing $p_t$. One can also note that the two calculations are
consistent with each other up to $p_t \simeq 3$ TeV, and only above this value
FONLL starts predicting a significantly smaller cross section.

\begin{figure}[t]
\center
\includegraphics[width=0.5\textwidth]{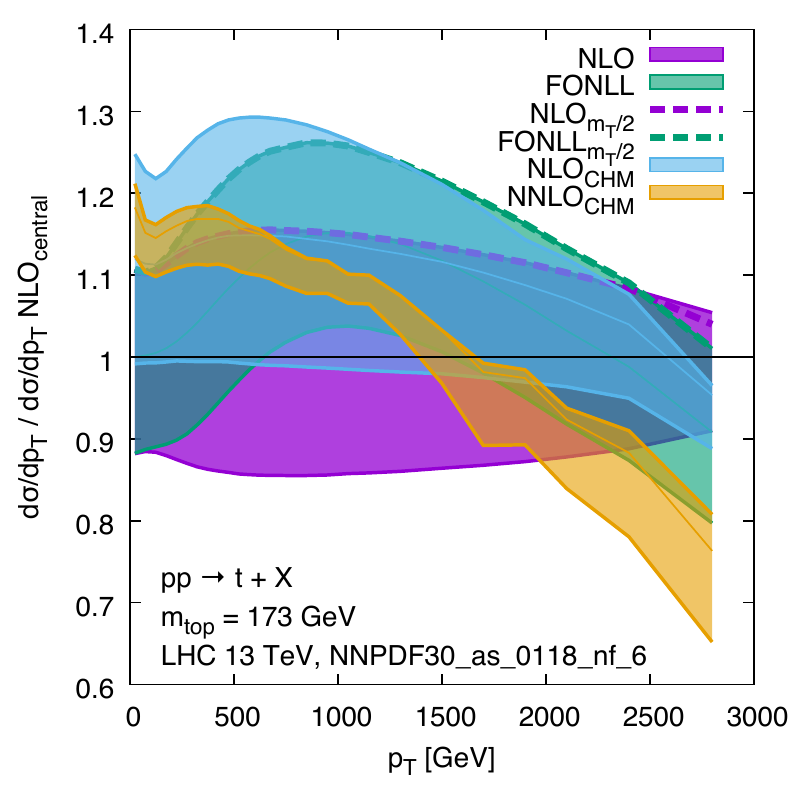}
\caption{\label{fig:comp} Comparison of FONLL transverse momentum
distribution for top quark production at the LHC with other calculations.}
\end{figure}

Figure~\ref{fig:comp} shows further comparisons of the FONLL result with other
calculations, and most notably with the NNLO fixed-order result from Czakon, 
Heymes and Mitov (CHM)~\cite{Czakon:2016dgf}.
An FONLL curve obtained setting the factorisation and renormalisation scales to
$m_T/2 \equiv \sqrt{m^2+p_t^2}/2$, rather than the default $m_T$, is shown in
this plot because $m_T/2$ is the default choice for the CHM calculations. The
main take-away from this plot is probably that, at large $p_t$, the NNLO
calculations seems to reproduce the softer behaviour suggested by the
resummation in FONLL. Nevertheless, in the region where data are presently
available ($p_t < 1$~TeV) the various calculations are likely not
discriminable by the data.

\begin{figure}[th]
\includegraphics[width=0.5\textwidth]{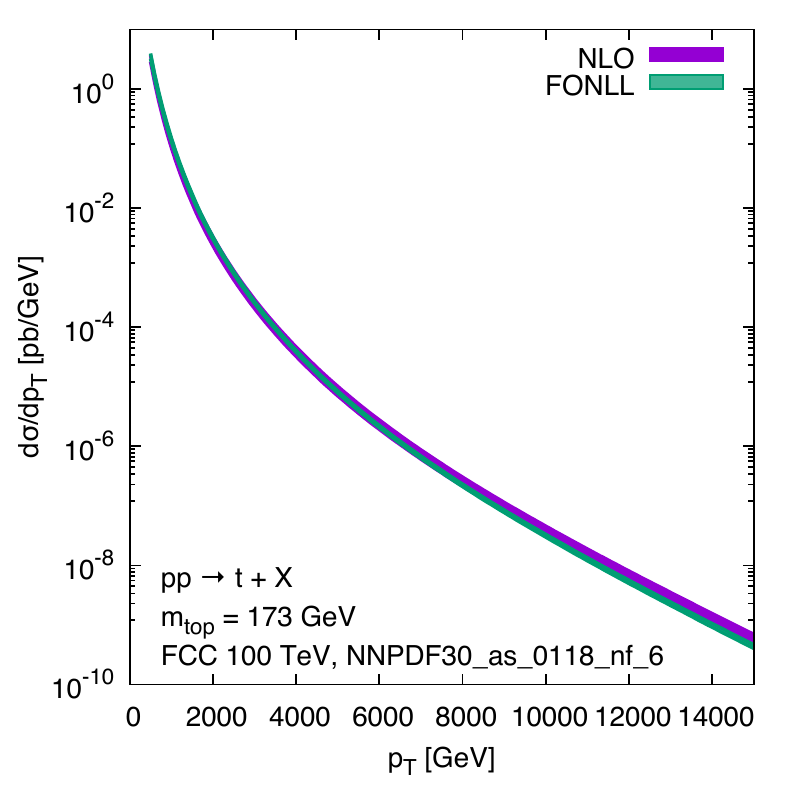}
\includegraphics[width=0.5\textwidth]{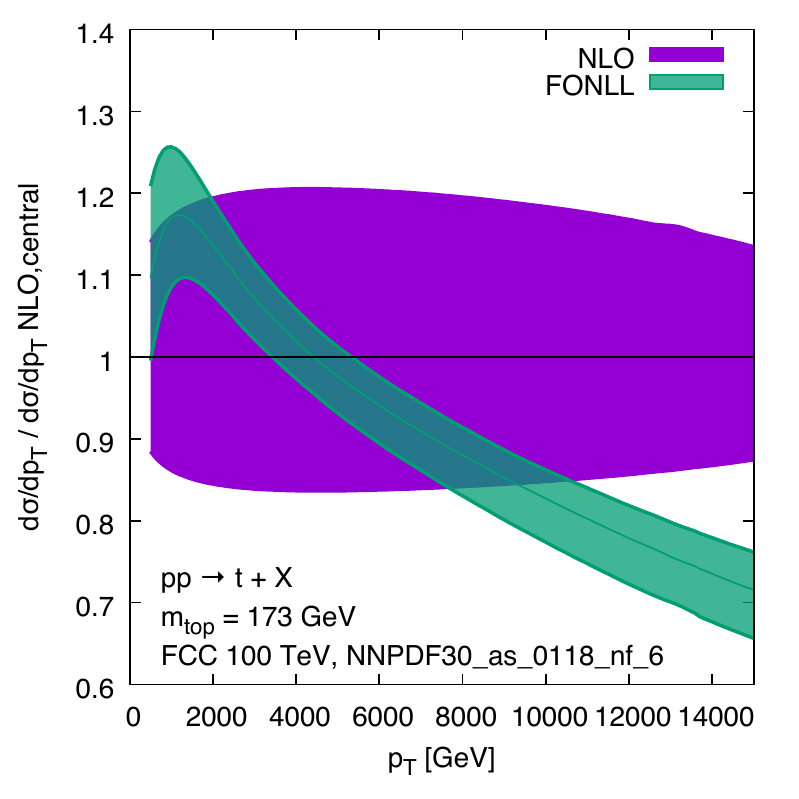}
\caption{\label{fig:fcc} Comparison of NLO and FONLL transverse momentum
distribution for top quark production at the FCC.}
\end{figure}

Finally, Figure~\ref{fig:fcc} shows the NLO and FONLL comparison for the
hypothetical future hadron collider at $\sqrt{S} = 100$~TeV, the FCC, and up to
extremely large top quark transverse momenta, $p_t = 15$~TeV. At these very
large
transverse momentum scales we observe a behaviour already seen for charm and
bottom at typical  LHC scales (i.e. order one hundred GeV), namely an FONLL band decisively narrower than the
NLO one, and a clearly softer transverse momentum distribution that leads to
predicting a smaller cross section at very large $p_t$.

\section{Conclusions}

A new version of the FONLL code, now implementing also top quark production and
delivering NLO+NLL accurate predictions for transverse momenta distributions,
has been presented.  Numerical results have been obtained for LHC and FCC
kinematics. They show that, at least in the transverse momentum region presently
explored by ATLAS and CMS, no significant difference  with respect to available
fixed order predictions is predicted by FONLL. The large transverse momentum
resummation of FONLL may instead become relevant if top is ever measured at
transverse momentum scales of several TeV.

The new version of the FONLL code will be made public in the near future,
accompanied by a paper~\cite{CDR} containing more extensive comparisons to other
calculations and to the data that are presently available.


\begin{thebibliography}{99}

\bibitem{Aaboud:2018eqg}
  M.~Aaboud {\it et al.} [ATLAS Collaboration],
  Phys.\ Rev.\ D {\bf 98} (2018) no.1,  012003
  [arXiv:1801.02052 [hep-ex]].

\bibitem{Sirunyan:2018wem}
  A.~M.~Sirunyan {\it et al.} [CMS Collaboration],
  Phys.\ Rev.\ D {\bf 97} (2018) no.11,  112003
  [arXiv:1803.08856 [hep-ex]].

\bibitem{Nason:1989zy}
  P.~Nason, S.~Dawson, R.~K.~Ellis,
  Nucl.\ Phys.\  {\bf B327 } (1989)  49-92;


\bibitem{Beenakker:1988bq}
W.~Beenakker, H.~Kuijf, W.~L.~van Neerven and J.~Smith,
Phys.\ Rev.\ D {\bf 40} (1989) 54.


\bibitem{Czakon:2015owf}
  M.~Czakon, D.~Heymes and A.~Mitov,
  Phys.\ Rev.\ Lett.\  {\bf 116} (2016) no.8,  082003
  [arXiv:1511.00549 [hep-ph]].
  
\bibitem{Czakon:2016ckf}
  M.~Czakon, P.~Fiedler, D.~Heymes and A.~Mitov,
  JHEP {\bf 1605} (2016) 034
  [arXiv:1601.05375 [hep-ph]].
  
\bibitem{Czakon:2016dgf}
  M.~Czakon, D.~Heymes and A.~Mitov,
  JHEP {\bf 1704} (2017) 071
  [arXiv:1606.03350 [hep-ph]].

\bibitem{Cacciari:1998it}
  M.~Cacciari, M.~Greco, P.~Nason,
  JHEP {\bf 9805 } (1998)  007 
  [hep-ph/9803400].
    
\bibitem{Cacciari:1993mq}
  M.~Cacciari and M.~Greco,
  Nucl.\ Phys.\ B {\bf 421} (1994) 530
  [hep-ph/9311260].

\bibitem{CDR} M. Cacciari, F. Dreyer and E. Re, in preparation
  
\end{thebibliography}
\end{document}